\documentclass[letter]{aa}
\usepackage{natbib}
\usepackage{graphicx}
\usepackage{txfonts}

\begin{document}
\title{Planet--planet scattering in circumstellar gas disks} 
\subtitle{}

\titlerunning{Planet scattering in gas disks}

\author{F. Marzari
        \inst{1},
        C. Baruteau
       {\inst{2}, and}
        H. Scholl
        \inst{3}
        }

   \offprints{F. Marzari}

   \institute{
              Dipartimento di Fisica, University of Padova, Via Marzolo 8,
              35131 Padova, Italy\\
              \email{marzari@pd.infn.it}
         \and
             Astronomy and Astrophysics Department, University of California,
             Santa Cruz, CA 95064, USA\\
             \email{clement.baruteau@ucolick.org}
         \and
              Laboratoire Cassiop\'ee, Universit{\'e} de Nice Sophia Antipolis, CNRS, 
              Observatoire de la C\^ote d'Azur, B.P. 4229, F-06304 Nice Cedex, France\\
             \email{Hans.Scholl@oca.eu}
             }

   \date{Received XXX ; accepted XXX}

\abstract 
{Hydrodynamical simulations of
two giant planets embedded in a gaseous disk have shown that
in case of a smooth convergent migration they end up trapped
into a mean motion resonance. 
These findings have led to
the conviction that the onset of dynamical instability causing
close encounters between the planets
can occur only after the
dissipation of the gas when the eccentricity damping is over.}
{
We show that a system of three giant planets
may undergo planet-planet scattering when the gaseous disk,
with density values comparable to that of the Minimum Mass Solar Nebula,
is still interacting with the planets.}
{The hydrodynamical code FARGO--2D--1D is used to model the evolution of
the disk and planets, modified to properly handle close encounters
between the massive bodies.}
{Our simulations predict
a variety of different outcomes of the scattering phase
which includes orbital exchange, planet
merging and scattering of a planet in a hyperbolic orbit.}
{This implies that
the final fate of a multiplanet system under the action of the disk torques
is not necessarily a packed resonant configuration.}


\maketitle

\section{Introduction}

\begin{figure*}[hp]
\begin{center}
\begin{tabular}{c c}
%\hskip -3 truecm
\hskip -2 truecm
\resizebox{80mm}{!}{\includegraphics[angle=-90]{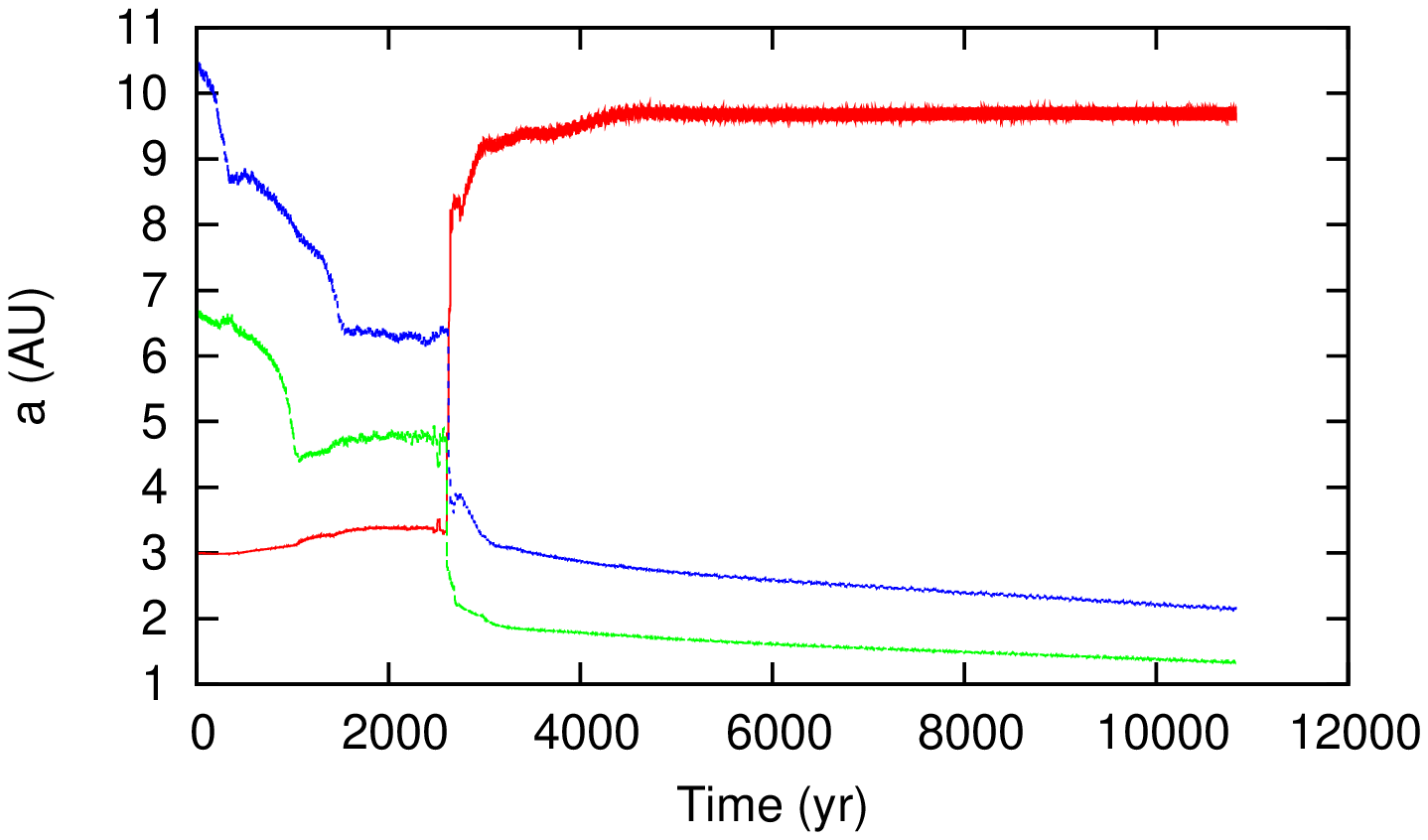}} %\hskip -6 truecm
\resizebox{80mm}{!}{\includegraphics[angle=-90]{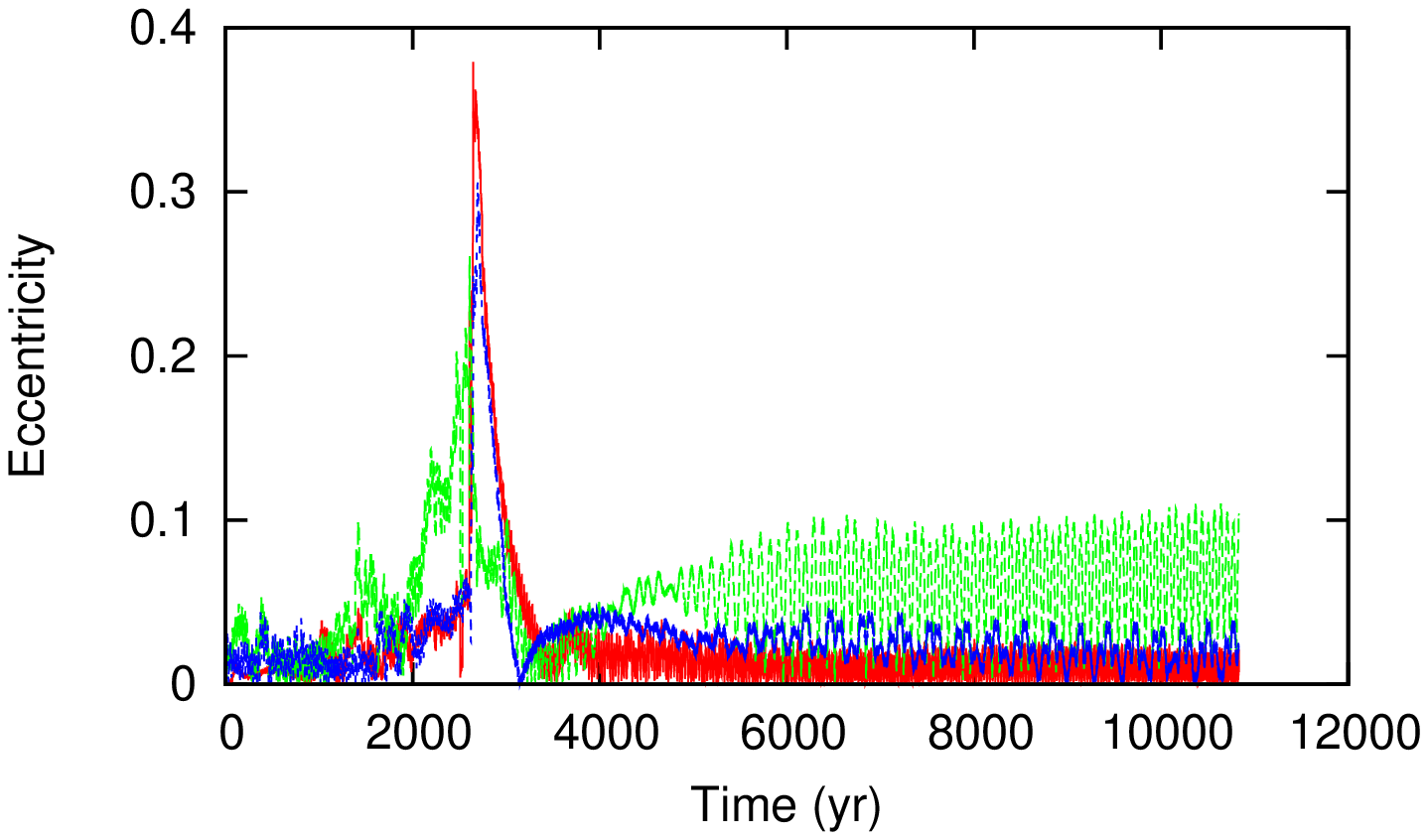}} \\
%\hskip -3 truecm
\hskip -2 truecm
\resizebox{80mm}{!}{\includegraphics[angle=-90]{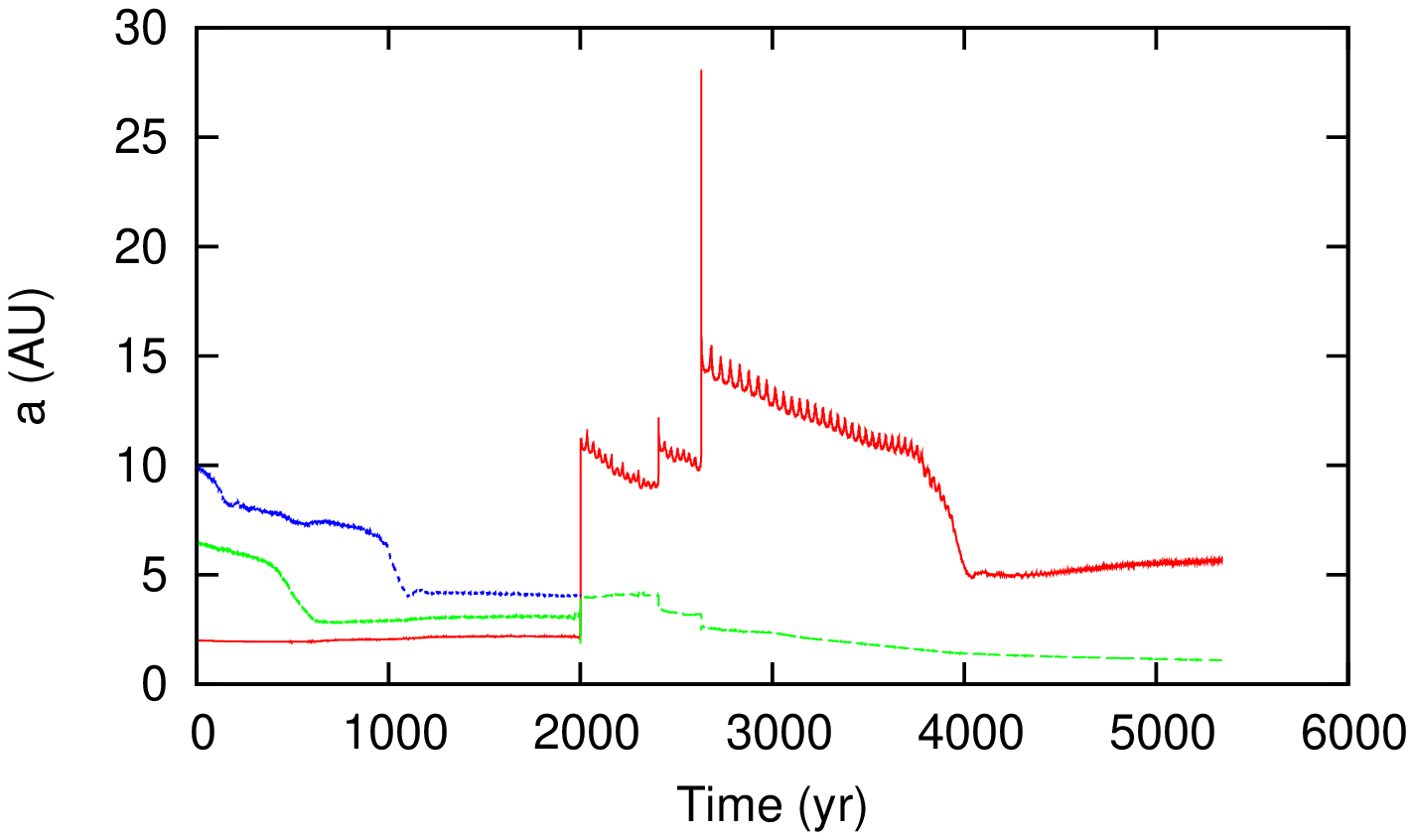}} %\hskip -6 truecm
\resizebox{80mm}{!}{\includegraphics[angle=-90]{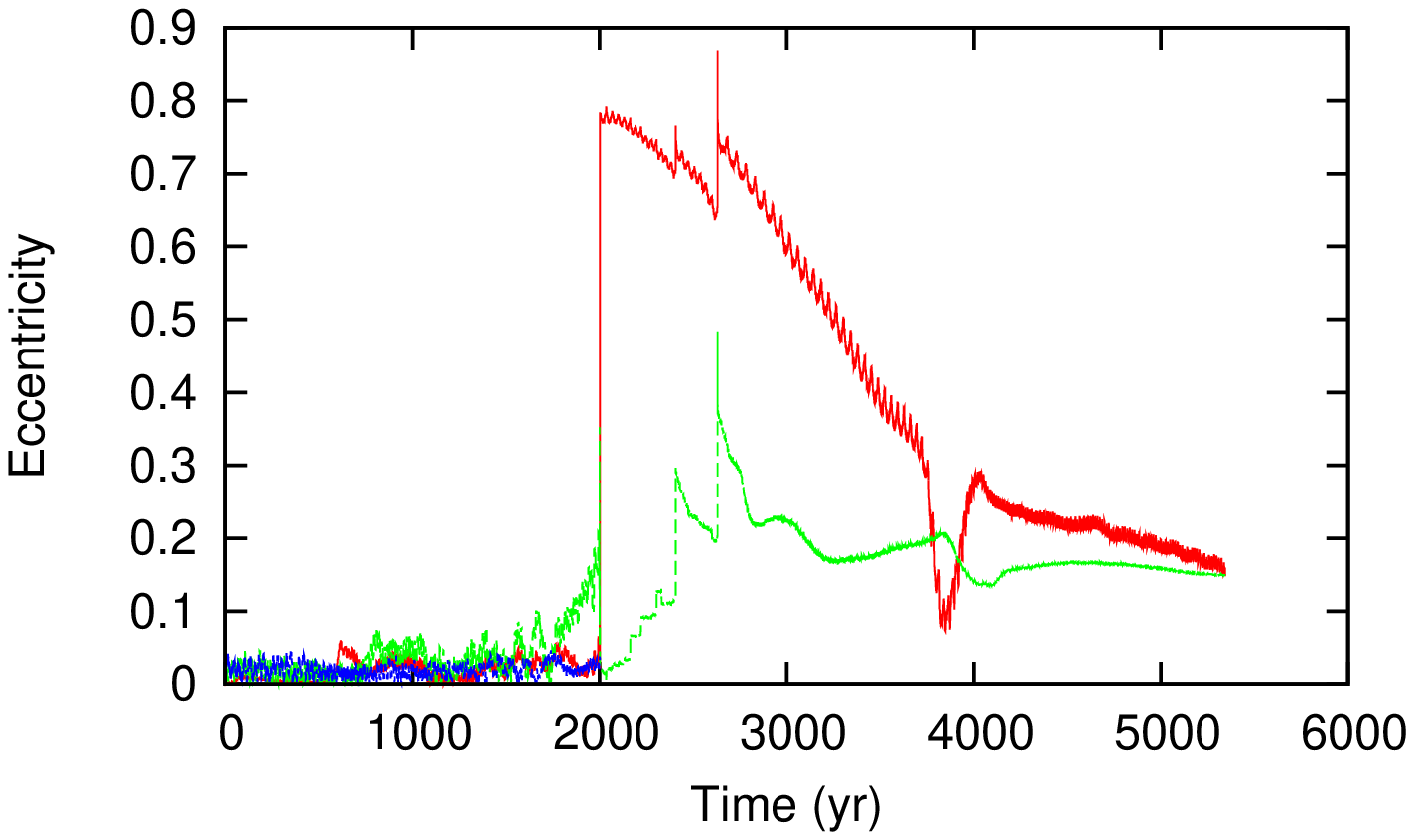}} \\
%\hskip -3 truecm
\hskip -2 truecm
\resizebox{80mm}{!}{\includegraphics[angle=-90]{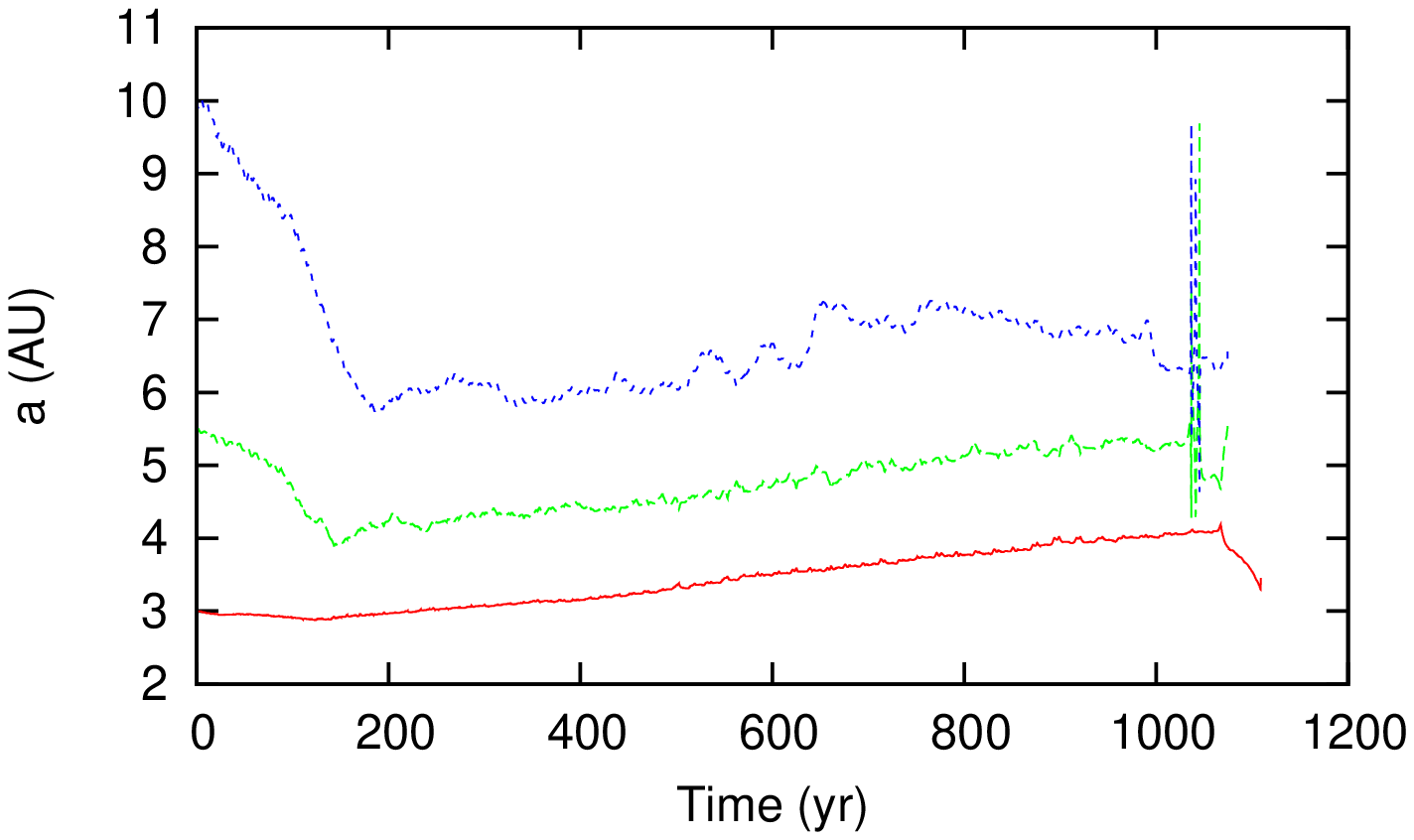}} %\hskip -6 truecm
\resizebox{80mm}{!}{\includegraphics[angle=-90]{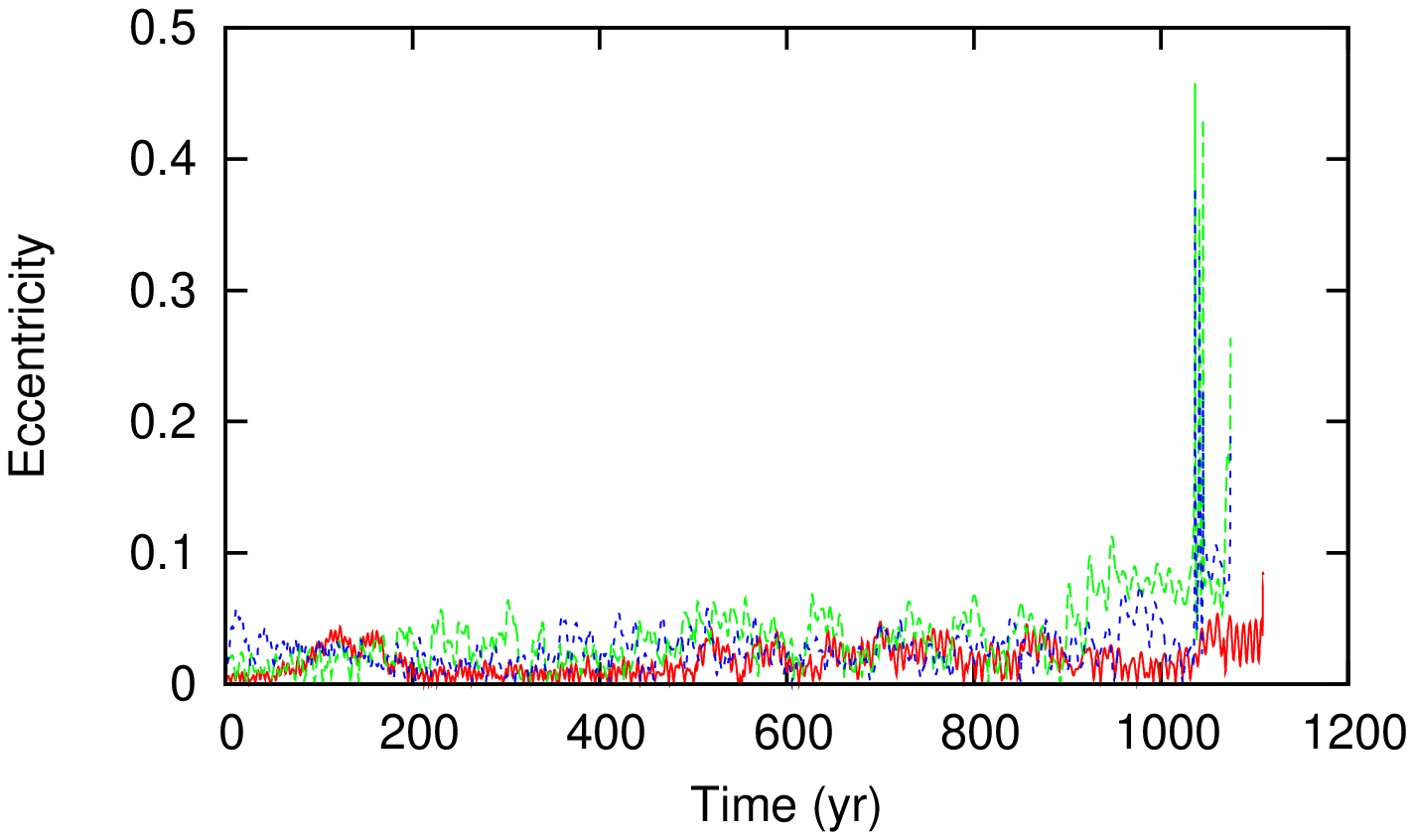}} \\
\end{tabular}
\caption[]{Orbital evolution of the three planets while embedded in the 
disk. The top plots show a model where $\Sigma_0 = 0.4 \Sigma_{\rm MMSN}$ 
and 
$a_1 = 3.0{\rm AU}, a_2 = 6.7{\rm AU}, a_3 =
10.5{\rm AU}$
The middle plots show a case with 
$\Sigma_0 = 0.65 \Sigma_{\rm MMSN}$ and 
$a_1 = 2.0{\rm AU}, a_2 = 6.5{\rm AU}, a_3 = 10.0{\rm AU}$
The bottom plots 
illustrate the evolution when $\Sigma_0 = \Sigma_{\rm MMSN}$ and
$a_1 = 3.0{\rm AU}, a_2 = 5.5{\rm AU}, a_3 = 10.0{\rm AU}$.
}
\label{f1}
\end{center}
\end{figure*}

A multi--planet system made of Jupiter-sized bodies
may reach completion when the planets are still
under the influence of the disk.  In this scenario the mutual gravitational
perturbations compete with the gravitational torques exerted on the planets
by the disk. These torques  cause planet migration which is
a strong source of mobility to the system and
possibly causes the planets to come closer to each other.
The mutual approach is
supposed to end  when a pair of planets is captured in a mean motion resonance
(e.g. the 5:3, 3:2, 3:1 or 2:1).
The most extensively studied scenario
is that of two jovian mass planets embedded in a
viscously evolving disk.  \cite{kpb, klm}, \cite{ms} and \cite{pn} have shown 
via self--consistent disk--planet
hydrodynamical simulations that the two planets become trapped into a resonance
and create a cavity encircling their orbits.
This resonant configuration appears stable in
these simulations which argues in favor of a scenario where the resonance locking
should be broken only after gas dissipation possibly by the end of the
gas damping or by planetesimal scattering.

Damped N--body models have also been used to study the
evolution of two planet systems where
the disk--planet interaction is modeled as an external
dissipative force. \cite{lp}, \cite{bmf} and \cite{ltr}
have shown that the resonance trapping is permanent
if a strong damping of eccentricity is adopted.
Even 
turbulence appears not to prevent the resonance capture \citep{rp} in
this scenario. The escape from resonance can occur 
only if the eccentricity damping term used in the model is relatively low 
and, after the resonance breaking, planets may have close encounters. 
This second approach
characterizes the outcome of the simulations by \cite{ma}, where two planets may
undergo planet--planet scattering after the escape from resonance
trapping. However, hydrodynamical simulations have so far never shown
such a behavior with the planets assumed to remain in resonance until
the dissipation of the gas. The gas damping appears to be dominating over
the mutual gravitational perturbations even for Jupiter size planets,
which supports the idea that the preferred outcome of the evolution
of a multiplanet system
under the tidal action of the disk is a packed resonant system.

A system of three giant planets
has proved to be dynamically more prone to instability and to develop crossing
orbits \citep{wm,cfmr} compared to the case of two planets.
The mutual gravitational perturbations are stronger and may more easily overcome 
the disk--damping effect. The observed systems Ups And, 55 Cnc, HIP 14810, HD74156,
HD160691 and HD 37124 also show that it is indeed possible to
form more than two giant planets around solar type stars.
For this reason we selected a
scenario where three giant planets with masses close to that  of
Jupiter  form in the disk and evolve under planet-disk
and planet-planet interactions. We will show that the chaotic evolution, 
which is characterized by mutual close encounters between (giant) planets,
is favored by planet migration. Our hydrodynamical simulations show 
that most of the time the planets are quickly captured in a three-body resonance
during their migration,  despite their rapid drift rate,
but this may become unstable. The growth in eccentricity
leads the planets into crossing orbits, and they begin to 
evolve under the effects of mutual approaches.  The 
following chaotic phase ends up either with the 
ejection of a planet, planet--merging or orbital reconfiguration. 

\section{Numerical set--up}

The evolution of the planets and disk was modeled with
the  hydrodynamical
code FARGO--2D--1D  \citep{cmm} where
the 
numerical integrator computing the planet orbits was updated 
with a variable stepsize so that mutual close encounters between the 
planets could be properly modeled. 
The 2D grid, which resolves the local disk-planet interactions,
is surrounded by a 1D (non azimuthally resolved)
grid that allowed us to simulate the disk's global evolution \citep{cmm}. 
The 2D grid used in the simulations to represent the disk has a
logarithmic radial spacing extending
from r=0.5 to r=15 AU. It is characterized by
$N_{\rm r}=512$ elementary rings, and $N_{\rm s}=256$ sectors.  The 1D grid
spans a radial distance from 15 AU outwards till 50 AU.
The disk--aspect ratio (H/r) was set uniformly to 5\%
and constant in time because the disk was assumed to be locally isothermal.
The gas kinematic viscosity is $\nu=10^{-5}$ in code--normalized
units, which corresponds to an alpha parameter of
$4\times 10^{-3}$. The gas density profile declines as $r^{-1/2}$ and the
value $\Sigma_0$ at 1 AU was varied in the simulations, ranging from 
0.25 to 1 $\Sigma_{\rm MMSN}$, the density at 1 AU predicted for the
Minimum Mass Solar Nebula. In the simulations presented here
the three planets have masses equal to 1.5, 0.9 and 1.2 Jupiter masses,
while the initial semi-major axes are selected to be between 2 and 10 AU. 
They are all
started on circular orbits and do not
accrete mass from the disk.

\section{Results}  

Figure 1 illustrates the different behaviors we observed in our simulations. 
After an initial fast migration, the three giant planets become
locked into a three--body mean motion resonance. However, the resonant 
trapping is not stable and the growth in eccentricity leads to 
planet--crossing orbits, which in turn start the chaotic phase. In the top plots we show the 
semi-major axis and eccentricity evolution of the three 
planets initially set on orbits with $a_1 = 3.0{\rm AU}, a_2 = 6.7{\rm AU}, a_3 =
10.5{\rm AU}$ with $\Sigma_0 = 0.4 \Sigma_{\rm MMSN}$. 
After repeated mutual encounters, an orbital exchange 
occurs in which
innermost planet becomes the outermost one. 
In the middle plots, a case of planet merging is displayed,
where the two surviving planets are left on mildly eccentric orbits.
In this model the initial disk density is $\Sigma_0 = 0.65 \Sigma_{\rm MMSN}$ 
and the planets are started at $a_1 = 2.0{\rm AU}, a_2 = 6.5{\rm AU}, 
a_3 = 10.0{\rm AU}$.
Very large eccentricities
are achieved during the chaotic phase, but they are partly 
damped by the disk after the merging event. 
It is interesting to note that in this second case, after the
planet--planet scattering, the two planets left on eccentric
orbits are surrounded with eccentric gaps. This is illustrated
in Fig. 2.
The bottom plots refer to a
simulation where two planets merge and are ejected out of the system
on a hyperbolic orbit ($\Sigma_0 = \Sigma_{\rm MMSN}$, $a_1 = 3.0{\rm AU}, 
a_2 = 5.5{\rm AU}, a_3 = 10.0{\rm AU}$). 

Different set ups were tested to check the
robustness of our findings. In one series of simulations, the mass of
each planet grows from 0 to the final value on a timescale ranging
from 10 to 1000 yrs,  the latter being 
comparable with the rapid gas--infall timescale typical 
of the final stage of a giant planet growth.  
In another series, planets are first held on fixed
circular orbits until they form gaps, after what they are let to evolve
under disk-planet and planet-planet interactions. Very similar
behaviors are observed in all these cases. The mutual 
gravitational interactions within a system of three planets
force some eccentricity, which is not fully damped 
by the disk.  In addition, the wakes excited by each 
planet tamper with the gap--cleaning process of the other
planets. The two outermost planets in our simulations do not 
undergo slow type II migration. They experience strong torques from 
their co--orbital region which is not fully cleared. The fast migration 
\citep[runaway migration,][]{mp03} triggered by corotation torques 
leads the planets close enough to be trapped into mean--motion resonance. 
But this is not stable and planet--planet scattering follows.

\begin{figure}
%\begin{figure*}[hp]
%\begin{center}
%\begin{tabular}{c}
%\hskip -3 truecm
\resizebox{\hsize}{!}{\includegraphics[angle=0]{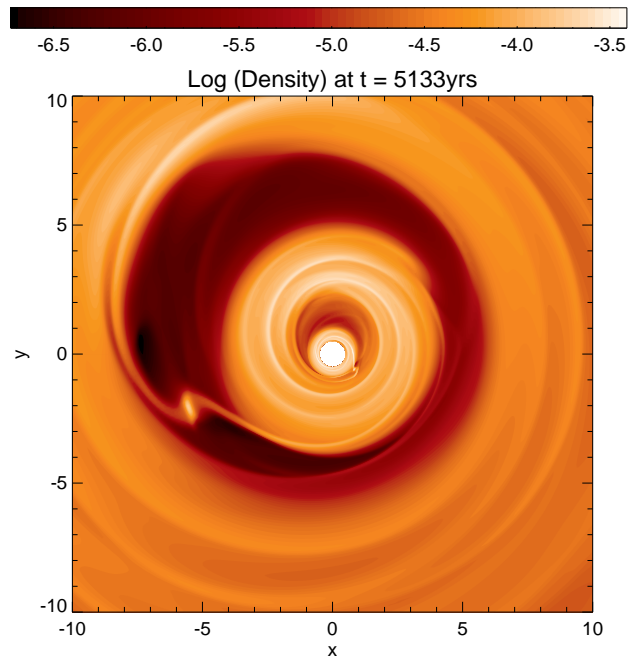}}
%\end{tabular}
\caption[]{Disk surface density distribution after the 
planet--planet scattering phase shown in Fig. 1 middle plots. 
The two planets have developed eccentric gaps. 
}
\label{f2}
%\end{center}
%\end{figure*}
\end{figure}

\section{Discussion and conclusions}

 Although there are known two-planet
systems in simple mean-motion resonances, there are also many systems that are
either multiple and non-resonant, or that display evidence for scattering that
would be suppressed if resonant capture was a ubiquitous process.
We show that resonant capture is more easily evaded in a
three-planet as opposed to a two-planet system, leading to a phase
of planet--planet scattering.
A major implication of planet--planet scattering in the
gas disk is the significant re--arrangement of planet orbits and
eccentricity excitation, which may affect the subsequent evolution of
the planets.  As a consequence, the final mass distribution of planets
as a function of the semi-major axis cannot be representative of the
formation process. In addition, a closely packed planetary system
where the multiple planets are all locked in mean--motion resonances is
not the preferred outcome of planet migration, as previously argued.
The chaotic phase may lead to different orbital configurations
where the planets can be well separated into semi-major axis and stable
even after the gas dissipation. Lower eccentricities are
expected as the final state of the planet evolution because damping can
occur after the  dynamically active phase \citep{moe}. 
However, it may not
be the case for corotation resonance saturation. According
to  \cite{gs, ol} a planet may experience a net growth of eccentricity via
disk--planet interaction if its eccentricity is not small to begin
with. This would be the case for a planet at the end of the chaotic
phase.  

Additional simulations are undergoing in order to perform 
a deeper exploration of the parameter space to 
identify the initial configurations of the planets and disk that lead to 
planet--planet scattering rather than to
a stable resonant trapping. 

\begin{acknowledgements}
We thank Phil Armitage for his useful comments and suggestions 
as referee of the paper. 
\end{acknowledgements}

\end{document}